\documentclass[twocolumn,twoside]{IEEEtran}
\usepackage{amsmath,amssymb,amsthm,wasysym,epsfig,color,subfigure,graphicx}
\usepackage{enumerate,url,algpseudocode,algorithm,balance,url,bm,nicefrac,caption,subfloat}

\newtheorem{proposition}{Proposition}

\theoremstyle{remark}\newtheorem{remark}{Remark}
\allowdisplaybreaks

\newcommand{\minimize}{{\rm minimize}}

\def\ccalH{{\ensuremath{\mathcal H}}}
\def\ccalT{{\ensuremath{\mathcal T}}}

\makeatletter

\begin{document}
\title{ Robust and Scalable Power System State\\ Estimation via
 Composite Optimization}

\author{
Gang Wang,~\IEEEmembership{Member,~IEEE},
	Georgios B. Giannakis,~\IEEEmembership{Fellow,~IEEE},
	and Jie Chen,  \IEEEmembership{Fellow,~IEEE}

\thanks{
The work of G. Wang and G. B. Giannakis was supported in part by the National Science Foundation under Grants 1514056, 1505970, and 1711471. The work of J. Chen was supported in part by the National Natural Science Foundation of China (NSFC) under Grants 61621063, 61522303, by the Projects of Major International (Regional) Joint Research Program NSFC under Grant 61720106011, and by the Program for Changjiang Scholars and Innovative Research Team in University (IRT1208).

G. Wang and G. B. Giannakis are with the Digital Technology Center and the Electrical and Computer Engineering Department, University of Minnesota, Minneapolis, MN 55455, USA (email: gangwang@umn.edu, georgios@umn.edu).
J. Chen is with the State Key Lab of Intelligent Control and Decision of Complex Systems and School of Automation, Beijing Institute of Technology, Beijing 100081, China, and also with Tongji University, Shanghai 200072, China (email: chenjie@bit.edu.cn).
}


}

\markboth{IEEE TRANSACTIONS ON SMART GRID (To appear, 2019)}{}

\maketitle

\begin{abstract}

In today's cyber-enabled smart grids, 
high penetration of uncertain renewables, 
purposeful manipulation of meter readings, 
and the need for wide-area situational awareness, call for fast, accurate, and robust power system state estimation. 
The least-absolute-value (LAV) estimator is known for its robustness relative to the weighted least-squares (WLS) one. 
However, due to nonconvexity and nonsmoothness,
existing LAV solvers based on linear programming 
are typically slow, hence inadequate for real-time system monitoring.   
This paper develops two novel algorithms for efficient LAV estimation, which draw from recent advances in composite optimization. 
The first is a deterministic linear proximal scheme that handles a sequence of ($5\sim 10$ in general) convex quadratic problems, each efficiently solvable either via off-the-shelf toolboxes or through the alternating direction method of multipliers. Leveraging the sparse connectivity inherent to power networks, the second scheme is stochastic, and updates only \emph{a few} entries of the complex voltage state vector per iteration. In particular, when voltage magnitude and (re)active power flow measurements are used only, this number reduces to one or two, \emph{regardless of} the number of buses in the network. This computational complexity evidently scales well to large-size power systems.  
Furthermore, by carefully \emph{mini-batching} the 
voltage and power flow
 measurements, accelerated implementation of the stochastic iterations becomes possible.
The developed algorithms are numerically evaluated using a variety of benchmark power networks. 
Simulated tests corroborate that improved robustness can be attained at comparable or markedly reduced computation times for medium- or large-size networks 
relative to existing alternatives.

\end{abstract}

\begin{keywords} SCADA measurements, nonlinear AC estimation, cyberattacks, alternating direction method of multipliers, prox-linear algorithms
\end{keywords}

\section{Introduction}\label{sec:intro}

The North American electric grid, the largest machine on earth, is recognized as the greatest engineering achievement of the $20$th century \cite{nae-report}: thousands of miles of transmission lines and millions of miles of distribution lines, linking thousands of power plants to millions of factories and homes.
 Accurately monitoring the grid's operating condition is critical to several control and optimization tasks, including optimal power flow, reliability analysis, attack detection, and future network expansion planning \cite{AburExpositoBook,dsse2018wgcs}.

To enable grid-wide monitoring, power system engineers in the $1960$s pursued voltages at critical buses based on readings  collected from current and potential transformers. But the power flow equations were never feasible due to timing and modeling inaccuracies.
In a seminal work \cite{Schweppe70}, the statistical foundation was laid for power system state estimation (PSSE), whose central task is to obtain the voltage magnitude and angle information at all buses given network parameters and measurements acquired from across the grid. Since then, there have been numerous PSSE contributions; 
see \cite{dsse2018wgcs} for a recent review of PSSE advances, some of which 
 are outlined below. 

Power grids are primarily monitored by supervisory control and data acquisition (SCADA) systems. 
Parameter uncertainty, instrument mis-calibration, and unmonitored topology changes can however, yield grossly corrupted SCADA measurements (a.k.a.
`bad data') \cite{1971baddata}. Designed for functionality and efficiency with little attention paid to security, today's SCADA systems are vulnerable to cyberattacks \cite{2016cybersecurity}. Bad data also come in the form of purposeful manipulation of smart meter readings, as asserted by the first hacker-caused power outage: the 2015 Ukraine blackout  \cite{ukraine}.
Any of these events can happen which will cause a given data collection to be much more inaccurate than is assumed by popular mathematical models. Efficient robust PSSE approaches against cyber threats are thus well motivated in the smart grid context \cite{scpse}.

Commonly used PSSE criteria include the weighted least-squares (WLS) and the least-absolute value (LAV) \cite{2012conejo}. Other enhanced estimators for robustness consist of the Schweppe-Huber generalized M-estimator \cite{mili1996robust}, \cite{tps2018zm}, \cite{tps2018zmp}, as well as the least-median and the least-trimmed squares state estimators \cite{mili1996robust}.   
The WLS criterion would coincide with the maximum likelihood criterion when additive white Gaussian noise is assumed. 
 Unfortunately, WLS comes with a number of limitations. First, obtaining the WLS estimate based on SCADA measurements amounts to minimizing a nonconvex quartic polynomial. As a result, the quality of the resultant iterative estimators relies heavily on the initialization (see justifications in e.g., \cite{taf,raf}). Furthermore, the convergence of Gauss-Newton iterations for nonconvex objectives is hardly guaranteed in general \cite{Be99}. As least as important, WLS estimators are sensitive to bad data \cite{1971baddata}. They may yield very poor estimates in the presence of outliers. 
 These issues were somewhat mitigated by incorporating the largest normalized residual (LNR) test for bad data removal \cite{1971baddata}, or, via reformulating the (possibly regularized) WLS into a semidefinite program (SDP) via convex relaxation \cite{jstsp2014zhu}, 
 \cite{psse2016zhang}. The former alternates between the LNR test and the estimation,
 while the latter solves SDPs.
The least-median-squares and the least-trimmed-squares estimators have provably improved performance under certain conditions \cite{Mili94}. 
Unfortunately, their computational complexities and storage requirements scale unfavorably with the number of buses in the network \cite{dsse2018wgcs}.

On the other hand, LAV estimators simultaneously identify and reject bad data while acquiring an accurate estimate of the state \cite{1982baddata}.
Recent research efforts have focused on dealing with the nonconvexity and nonsmoothness in LAV estimation.
Upon linearizing the nonlinear measurement functions at the most recent iterate, a series of linear programs was solved \cite{1982baddata}. 
Techniques for improving the linear programming by exploiting the system's structure \cite{1991fast}, or via iterative reweighting \cite{gtd2003jabr} have also been reported. LAV estimation based only on PMU data was studied \cite{2014abur,tsg2017xa}, in which a strategic scaling was suggested to eliminate the effect of leverage measurements \cite{Mili94,mili1996robust}. Despite these efforts, LAV estimators have not been widely employed yet in today's power networks due mostly to their computational inefficiency \cite{2014abur}. 

The LAV-based PSSE is revisited in this work from the viewpoint of composite optimization  \cite{1985co}, \cite{mp2016pl},
 which considers minimizing functions $f(\mathbf{v})=h(c(\mathbf{v}))$ that are compositions of 
a convex function $h$, and a smooth vector function $c$. 
Two novel proximal linear (prox-linear) procedures are developed based upon minimizing a sequence of convex quadratic subproblems.
The first deterministic LAV solver minimizes functions constructed from a linearized approximation to the original objective and a quadratic regularization, each efficiently implementable using either off-the-shelf solvers, or, the alternating direction method of multipliers (ADMM). The convergence of such deterministic prox-linear algorithms has been documented \cite{mp2016pl}, \cite{2016plconvergence}. 

The second LAV solver builds on a stochastic prox-linear algorithm, and has each iteration minimizing the summation of a linearized approximation to the LAV loss of a single measurement and the regularization term \cite{duchi2017stochastic,tcns2018wzgs}. 
Interestingly, each iteration of the stochastic LAV solver has a closed-form update. Thanks to the sparse connectivity inherent to power networks, this amounts to updating very few entries of the state vector. 
Moreover, even faster implementation of the stochastic solver is realized by means of judiciously mini-batching the measurements.

Bad leverage points may challenge, but the proposed prox-linear algorithms can be generalized to accommodate robust estimation formulations including the Huber estimation, Huber M-estimation, and the Schweppe-Huber generalized M-estimation \cite{mili1996robust}, \cite{tsp2010mili},  \cite{tps1999irls}, \cite{tps2018zm}, \cite{tps2018zmp}.
The novel algorithms were numerically tested using the IEEE $14$-, $118$-bus, and the PEGASE $9,241$-bus benchmark networks. Simulations corroborate their merits relative to the WLS-based Gauss-Newton iterations.

\emph{Outline.} Grid modeling  and problem formulation are given in Section \ref{sec:prel}. Upon reviewing the basics of composite optimization, Section \ref{sec:dlav} presents the deterministic LAV solver, followed by its stochastic alternative in Section \ref{sec:slav}. Extensive numerical tests are presented in Section \ref{sec:test}, while the paper is concluded in Section \ref{sec:conc}.  

\emph{Notation.} Matrices (column vectors) are denoted by upper- (lower-) case boldface letters.
Symbols $(\cdot)^{\!\ccalT}\!$ and $(\cdot)^{\!\ccalH}$ represent (Hermitian) transpose, and $\overline{(\cdot)}$ complex conjugate.
Sets are denoted using calligraphic letters.  Symbol ${\Re}(\cdot)$ (${\Im}(\cdot)$) takes the real (imaginary) part of a complex number. Operator ${\rm dg}(\mathbf{x}_i)$ defines a diagonal matrix holding entries of $\mathbf{x}_i$ on its diagonal, while $[\mathbf{x}_i]_{1\le i\le N}$ returns a matrix with $\mathbf{x}_i^\ccalH$ being its $i$-th row.

\section{Grid Modeling and Problem Formulation}
\label{sec:prel}

An electric grid having $N$ buses and $L$ lines is modeled as a graph $\mathcal{G}=(\mathcal{N},\,\mathcal{E})$, whose nodes $\mathcal{N}:=\{1,\,2,\,\ldots,\,N\}$ correspond to buses and whose edges $\mathcal{E}:=\{(n,\,n')\}\subseteq\mathcal{N}\times \mathcal{N}$ correspond to lines.
The complex voltage per bus $n\in\mathcal{N}$ is expressed in rectangular coordinates as $v_n= \Re(v_{n})+j\Im(v_{n})$, with all nodal voltages forming the vector $\mathbf{v}:=[v_1~\cdots~v_N]^\ccalH\in\mathbb{C}^N$.

The voltage magnitude square $V_n:=|v_n|^2=\Re^2(v_{n})+\Im^2(v_{n})$ can be compactly expressed as a quadratic function of $\mathbf{v}$, namely
\begin{equation}\label{eq:squa}
V_n=\mathbf{v}^\ccalH\mathbf{H}_{n}^V\mathbf{v},\quad {\rm with~~}  \mathbf{H}_{n}^V:=\mathbf{e}_n\mathbf{e}_n^\ccalT
\end{equation}
 where $\mathbf{e}_n$ denotes the $n$-th canonical vector in $\mathbb{R}^N$. To express power injections as functions of $\mathbf{v}$, introduce the so-termed bus admittance matrix $\mathbf{Y}=\mathbf{G}+j\mathbf{B}\in\mathbb{C}^N$ \cite{AburExpositoBook}. In rectangular coordinates, the active and reactive power injections $p_n$ and $q_n$ at bus $n$ are given by
 \begin{align}
p_n&=\Re(v_{n})\sum_{n'=1}^N[\Re(v_{n'})G_{nn'}-\Im(v_{n})B_{nn'}]\nonumber\\ 
&\qquad +\Im(v_{n})\sum_{n'=1}^N[\Im(v_{n'})G_{nn'}+\Re(v_{n'})B_{nn'}]\\
q_n&=\Im(v_{n})\sum_{n'=1}^N[\Re(v_{n'})G_{nn'}-\Im(v_{n})B_{nn'}]\nonumber\\ &\qquad -\Re(v_{n})\sum_{n'=1}^N[\Im(v_{n'})G_{nn'}+\Re(v_{n'})B_{nn'}]
 \end{align}
 which admits a compact representation as
 \begin{subequations}\label{eq:inje}
 	\begin{align}
 	p_n&=\mathbf{v}^\ccalH\mathbf{H}_{n}^p\mathbf{v},{\rm ~~with~~}\mathbf{H}_{n}^p:=\frac{\mathbf{Y}^\ccalH\mathbf{e}_n\mathbf{e}_n^\ccalT+\mathbf{e}_n\mathbf{e}_n^\ccalT\mathbf{Y}}{2}
 	\label{eq:ainj}\\
 	q_n&=\mathbf{v}^\ccalH\mathbf{H}_{n}^q\mathbf{v}, {\rm ~~with~~}\mathbf{H}_{n}^q:=\frac{\mathbf{Y}^\ccalH\mathbf{e}_n\mathbf{e}_n^\ccalT-\mathbf{e}_n\mathbf{e}_n^\ccalT\mathbf{Y}}{2j}\label{eq:rinj}.
 	\end{align}
 \end{subequations}

Recognize that the line current from bus $n$ to $n'$ at the `from' end obeys $I_{nn'}=\mathbf{e}_{nn'}^\ccalT\mathbf{i}_{f}=\mathbf{e}_{nn'}^\ccalT\mathbf{Y}_f\mathbf{v}$, where $\mathbf{i}_f\in\mathbb{C}^{|\mathcal{E}|}$ collects all line currents, and $\mathbf{Y}_f\in\mathbb{C}^{|\mathcal{E}|\times N}$ relates the bus voltages to all line currents at the `from' (sending) end.
Ohm's and Kirchhoff's laws assert that the `from-end' power flow over line $(n,\,n')$ can be expressed as $\overline{S}_{nn'}^f=P_{nn'}^f-jQ_{nn'}^f=\overline{v_n} i_{nn'}^f=(\mathbf{v}^\ccalH\mathbf{e}_n)(\mathbf{e}_{nn'}^\ccalT\mathbf{i}_f)=\mathbf{v}^\ccalH\mathbf{e}_n\mathbf{e}_{nn'}^\ccalT\mathbf{Y}_f\mathbf{v}$, yielding
\begin{subequations}
	\label{eq:flow}
	\begin{align}
	P_{nn'}^f&=\mathbf{v}^\ccalH\mathbf{H}_{nn'}^P\mathbf{v},
	{\rm ~with~}\mathbf{H}_{nn'}^P:=\frac{\mathbf{Y}_f^\ccalH\mathbf{e}_{nn'}\mathbf{e}_n^\ccalT+\mathbf{e}_n\mathbf{e}_{nn'}^\ccalT\mathbf{Y}_f}{2}\label{eq:aflo}\\
		Q_{nn'}^f&=\mathbf{v}^\ccalH\mathbf{H}_{nn'}^Q\mathbf{v},{\rm ~with~}\mathbf{H}_{nn'}^Q:=\frac{\mathbf{Y}_f^\ccalH\mathbf{e}_{nn'}\mathbf{e}_n^\ccalT-\mathbf{e}_n\mathbf{e}_{nn'}^\ccalT\mathbf{Y}_f}{2j}.\label{eq:rflo}
	\end{align}
\end{subequations}
The active and reactive power flows measured at the `to' (receiving) ends $P_{nn'}^t$ and $Q_{nn'}^t$ can be written symmetrically to $P_{nn'}^f$ and $Q_{nn'}^f$; and hence, they are omitted here for brevity.

Given line parameters collected in $\mathbf{Y}$ and $\mathbf{Y}_f$, all SCADA measurements including squared voltage magnitudes as well as active and reactive power injections and flows can be expressed as quadratic functions of the voltages  $\mathbf{v}\in\mathbb{C}^N$. This justifies why $\mathbf{v}$ is referred to as the system state. If $\mathcal{S}_{V}$, $\mathcal{S}_{p}$, $\mathcal{S}_{q}$, $\mathcal{S}_{P}^f$, $\mathcal{S}_{Q}^f$, $\mathcal{S}_{P}^t$, and $\mathcal{S}_{Q}^t$ signify the smart meter locations of the corresponding type, we have available the following (possibly noisy or even corrupted) measurements:
$\{\check{V}_n\}_{n\in\mathcal{S}_{V}}$, $\{\check{p}_n\}_{n\in\mathcal{S}_{p}}$, $\{\check{q}_n\}_{n\in\mathcal{S}_{q}}$,  $\{\check{P}_{nn'}^f\}_{(n,n')\in\mathcal{S}_{P}^f}$, $\{\check{Q}_{nn'}^f\}_{(n,n')\in\mathcal{S}_{Q}^f}$,
$\{\check{P}_{nn'}^t\}_{(n,n')\in\mathcal{S}_{P}^t}$, and  $\{\check{Q}_{nn'}^t\}_{(n,n')\in\mathcal{S}_{Q}^t}$,
 henceforth concatenated in the vector $\mathbf{z}\in\mathbb{R}^{M}$, where $M$ denotes the total number of measurements.

In this paper, the following corruption model is considered \cite{duchi2017}: If $\{\xi_i\}\subseteq \mathbb{R}$ models an arbitrary attack (or outlier) sequence, given the measurement matrices $\{\mathbf{H}_m\}_{m=1}^M$ in \eqref{eq:squa}-\eqref{eq:aflo}, we observe for $1\le m\le M$ the samples
\begin{equation}\label{eq:meas}
z_m\approx\left\{\begin{array}{ll}
\mathbf{v}^\ccalH\mathbf{H}_m\mathbf{v}&{\rm if~} m\in\mathcal{I}^{nom}\\
\xi_m&{\rm if~} m\in\mathcal{I}^{out}
\end{array}
\right.
\end{equation}
where 
additive measurement noise can be included if $\approx$ is replaced by equality, and $\mathcal{I}^{nom},\,\mathcal{I}^{out}\subseteq \{1,2,\ldots,M\}$ collect the indices of nominal data and outliers, respectively. In other words, $\mathcal{I}^{out}$ is the set of meter indices that can be compromised. 
The indices in $\mathcal{I}^{out}$ are assumed chosen randomly from $\{1,2,\ldots,M\}$. Instrument failures occur at random, although the attack sequence $\{\xi_m\}$ may rely on $\{\mathbf{H}_m\}$ (even adversarially). 
Specifically, two models will be considered for the attacks.
\begin{enumerate}
	\item[\textbf{M1}] Matrices $\{\mathbf{H}_m\}_{m=1}^M$ are independent of $\{\xi_m\}_{m=1}^M$.
	\item[\textbf{M2}] Nominal measurement matrices $\{\mathbf{H}_m\}_{m\in\mathcal{I}^{nom}}$ are independent of  $\{\xi_m\}_{m\in\mathcal{I}^{out}}$.
\end{enumerate} 

It is worth highlighting that {M1} requires full independence between the corruption and measurements. That is, the attacker may only corrupt $\xi_m$ without knowing $\mathbf{H}_m$. On the contrary, {M2} allows completely arbitrary dependence between $\xi_m$ and $\mathbf{H}_m$ for $m\in\mathcal{I}^{out}$, which is natural as the type of corruption may also rely on the individual measurement $\mathbf{H}_m$ being taken.

Having elaborated on the system and corruption models, the PSSE problem can be stated as follows:  
Given matrices $\mathbf{Y}$, $\mathbf{Y}_f$, and the available measurements $\mathbf{z}\in\mathbb{R}^M$, with entries as in \eqref{eq:meas} obeying {M1} or {M2}, recover the voltage vector $\mathbf{v}\in\mathbb{C}^N$. The first attempt may be seeking the WLS estimate, or the ML one when assuming independent Gaussian noise \cite{Schweppe70}. 
It is known however that the WLS criterion is sensitive to outliers, and may yield very bad estimates even if there are few grossly corrupted measurements \cite{1971baddata}.
As is well documented in statistics and optimization, the $\ell_1$-based losses yield median-based estimators \cite{book2011huber}, and handle gross errors in the measurements $\mathbf{z}$ in a relatively benign way. Prompted by this, we will consider here minimizing the $\ell_1$ loss of the residuals, which leads to the so-called LAV estimate \cite{1982baddata}
\begin{equation}\label{eq:lav}
\underset{\mathbf{v}\in\mathbb{C}^N}{\text{minimize}}~~f(\mathbf{v}):=\frac{1}{M}\sum_{m=1}^M\left|\mathbf{v}^\ccalH\mathbf{H}_m\mathbf{v}-z_m\right|.
\end{equation}
Because of $\{\mathbf{v}^\ccalH\mathbf{H}_m\mathbf{v}\}_{m=1}^M$ and the absolute-value operation, the LAV objective in \eqref{eq:lav} is nonsmooth, nonconvex, and not even locally convex near the optima $\pm\mathbf{v}^\ast$. This is clear from the real-valued scalar case $f(v)=|v^\ast v-1|$, where $v\in \mathbb{R}$. A local analysis based on convexity and smoothness is thus impossible, and $f(\mathbf{v})$ is difficult to minimize. For this reason, Gauss-Newton is not applicable to minimize \eqref{eq:lav}. Nevertheless, the criterion $f(\mathbf{v})$ possesses several unique structural properties, which we exploit next to develop efficient algorithms.

{\color{black}
\begin{remark}\label{rmk:1}
	For an $N$-bus power system, most existing PSSE approaches have relied on optimizing over $(2N-1)$ real variables, which consist of either the polar or the rectangular components of the complex voltage phasors after excluding the angle or the imaginary part of the reference bus that is often set to $0$. 
	Nevertheless, when iterative algorithms are used, working directly with the $N$-dimensional complex voltage vector has in general lower complexity and computational burden than in the real case. This is due to the compact quadratic representations of all SCADA quantities in complex voltage phasors, namely the natural sparsity of quadratic measurement matrices in the unknown complex voltage phasor vector.
\end{remark}
}

\section{Deterministic Prox-Linear LAV Solver}\label{sec:dlav}

In this section, we will develop a deterministic solver of \eqref{eq:lav}. To that end, let us start rewriting the objective in \eqref{eq:lav} as 
\begin{equation}
	\underset{\mathbf{v}\in\mathbb{C}^N}{\text{minimize}}~~f(\mathbf{v}):=h(c(\mathbf{v}))
\end{equation}
the composition of a convex function $h:\mathbb{R}^M\to \mathbb{R}$, and a smooth vector function $c:\mathbb{C}^N\to \mathbb{R}^M$, a structure that is known to be amenable to efficient algorithms \cite{1985co}, \cite{mp2016pl}. It is clear that this general form subsumes \eqref{eq:lav} as a special case, for which we can take $h(\mathbf{u})=(1/M)\|\mathbf{u}\|_1$ and $c(\mathbf{v})=\left[\mathbf{v}^\ccalH\mathbf{H}_m\mathbf{v}-z_m\right]_{1\le m\le M}$. 
The compositional structure lends itself well to the proximal linear (prox-linear) algorithm, which is a variant of the Gauss-Newton iterations \cite{1985co}. Specifically, define close to a given $\mathbf{v}$ the local ``linearization'' of $f$ as 
\begin{equation}
	\label{eq:llin}
	f_{\mathbf{v}}(\mathbf{w}):=h\big(c(\mathbf{v})+\Re(\nabla^\ccalH c(\mathbf{v})(\mathbf{w}-\mathbf{v}))\big)
\end{equation}
where $\nabla c(\mathbf{v})\in\mathbb{C}^{N\times M}$ denotes the Jacobian matrix of $c$ at $\mathbf{v}$ based on Wirtinger derivatives for functions of complex-valued variables \cite[Appendix]{tsp2017wzgs}. In contrast to the nonconvex $f(\mathbf{v})$, function $f_{\mathbf{v}}(\mathbf{w})$ in \eqref{eq:llin} is convex in $\mathbf{w}$, which is the key behind the prox-linear method. Starting with some point $\mathbf{v}_0$, which can be the flat-voltage profile point, namely the all-one vector, construct the iteration
\begin{equation}
\label{eq:upda}
\mathbf{v}_{t+1}:=	\underset{\mathbf{v}\in\mathbb{C}^N}{\arg\min} \left\{f_{\mathbf{v}_t}(\mathbf{v})+\frac{1}{2\mu_t}\|\mathbf{v}-\mathbf{v}_{t}\|_2^2 \right\}
\end{equation}
where $\mu_t>0$ is a stepsize that can be fixed in advance, or be determined by a line search \cite{mp2016pl}.

Evidently, the subproblem \eqref{eq:upda} to be solved at every iteration of the prox-linear algorithm is convex, and can be handled by off-the-shelf solvers such as CVX \cite{2008cvx}. However, these interior-point based solvers may not scale well when $\{\mathbf{H}_m\}$ are large. For this reason, we derive next a more efficient iterative procedure using ADMM iterations \cite{Boyd10}, \cite{KeGi12}.

When specifying $f$ to be the LAV objective of \eqref{eq:lav}, the minimization in \eqref{eq:upda} becomes
\begin{equation}\label{eq:midd}
{\mathbf{v}}_{t+1}\!=\!\arg\underset{\mathbf{v}\in\mathbb{C}^N}{\min}\left\|\Re(\mathbf{A}_t(\mathbf{v}-\mathbf{v}_t))-\mathbf{c}_t\right\|_1\!+\!\frac{1}{2}\!\left\|\mathbf{v}-\mathbf{v}_t\right\|_2^2
\end{equation}
with coefficients given by
 \begin{subequations}\label{eq:dc}
 	\begin{align}
 	\mathbf{A}_t&:=\left[(2\mu_t/M)\mathbf{v}_t^\ccalH\mathbf{H}_m \right]_{1\le m\le M}\label{eq:dca}\\
 	\mathbf{c}_t&:=\left[(\mu_t/M)(z_m-\mathbf{v}_t^\ccalH\mathbf{H}_m\mathbf{v}_t)\right]_{1\le m\le M}\label{eq:dcc}.
 	\end{align}
 \end{subequations}
For brevity, let $\mathbf{w}:=\mathbf{v}-\mathbf{v}_t$, and rewrite \eqref{eq:midd} equivalently as a constrained optimization problem 
\begin{subequations}\label{eq:cons}
	\begin{align}
	\underset{\mathbf{u}\in\mathbb{C}^M,\;\mathbf{w}\in\mathbb{C}^N}{\minimize}~&~\left\|\Re(\mathbf{u})-\mathbf{c}_t\right\|_1+\frac{1}{2}\|\mathbf{w}\|_2^2\label{eq:cons1}\\
	{\rm subject~to\,}~&~\mathbf{A}_t\mathbf{w}=\mathbf{u}\label{eq:const2}.
	\end{align} 
\end{subequations}
To decouple constraints and also facilitate the implementation of ADMM, introduce an auxiliary copy $\tilde{\mathbf{u}}$ and $\tilde{\mathbf{w}}$ for $\mathbf{u}$ and $\mathbf{w}$ accordingly, and rewrite \eqref{eq:cons} into
\begin{subequations}\label{eq:admmp}
	\begin{align}
	\underset{\tilde{\mathbf{u}},\,\tilde{\mathbf{w}},\,\mathbf{u},\,\mathbf{w}}{\minimize}~&~\left\|\Re(\tilde{\mathbf{u}})-\mathbf{c}_t\right\|_1+\frac{1}{2}\left\|\tilde{\mathbf{w}}\right\|_2^2\label{eq:admmp1}\\
	{\rm subject~to}~&~\,\tilde{\mathbf{u}}=\mathbf{u},\;\tilde{\mathbf{w}}=\mathbf{w},\;
	\mathbf{A}_t\mathbf{w}=\mathbf{u}\label{eq:admmp2}.
	\end{align} 
\end{subequations}

Letting $\bm{\lambda}\in\mathbb{C}^N$ and $\bm{\nu}\in\mathbb{C}^M$ be the Lagrange multipliers corresponding to the $\mathbf{w}$- and $\mathbf{u}$-consensus constraints, respectively, the augmented Lagrangian after leaving out the last equality in \eqref{eq:admmp2} can be expressed as
\begin{align}
\mathcal{L}(\tilde{\mathbf{w}},\tilde{\mathbf{u}},\mathbf{w},\mathbf{u};\bm{\lambda},\bm{\nu})&:=\left\|\Re(\tilde{\mathbf{u}})-\mathbf{c}_t\right\|_1+\frac{1}{2}\!\left\|\tilde{\mathbf{w}}\right\|_2^2\nonumber\\
&\quad +\Re(\bm{\lambda}^\ccalH(\tilde{\mathbf{w}}-\mathbf{w}))+\Re(\bm{\nu}^\ccalH(\tilde{\mathbf{u}}-\mathbf{u}))
\nonumber\\
&\quad+\frac{\rho}{2}\left\|\tilde{\mathbf{w}}-\mathbf{w}\right\|_2^2+\frac{\rho}{2}\left\|\tilde{\mathbf{u}}-\mathbf{u}\right\|_2^2
\end{align}
where $\rho>0$ is a predefined step size. With $k\in\mathbb{N}$ denoting the iteration index for solving \eqref{eq:cons}, or equivalently \eqref{eq:upda}, 
ADMM cycles through the following recursions
\begin{subequations}
	\label{eq:admm}
	\begin{align}
	&	\tilde{\mathbf{w}}^{k+1}:=\!	\underset{\tilde{\mathbf{w}}}{\arg\min}\!\left\{\frac{1}{2}\|\tilde{\mathbf{w}}\|_2^2+\frac{\rho}{2}\left\|\tilde{\mathbf{w}}-(\mathbf{w}^k-\bm{\lambda}^k)\right\|_2^2	\right\}\label{eq:admm1}\\
	&	\tilde{\mathbf{u}}^{k+1}:=\!\underset{\tilde{\mathbf{u}}}{\arg\min}\!\left\{\frac{1}{2}\!\left\|\Re(\tilde{\mathbf{u}})-\mathbf{c}_t\right\|_1\!+\frac{\rho}{2}\left\|\tilde{\mathbf{u}}-\!(\mathbf{u}^k-\!\bm{\nu}^k)\right\|_2^2	\right\}\label{eq:admm2}\\
	& \left\{\mathbf{w}^{k+1},\mathbf{u}^{k+1}\right\}	:=\nonumber\\
	&\quad\underset{\mathbf{w},\;\mathbf{u}}{\arg\min}~ \left\|\mathbf{w}-(\tilde{\mathbf{w}}^{k+1}+\bm{\lambda}^{k})\right\|_2^2+\left\|\mathbf{u}-(\tilde{\mathbf{u}}^{k+1}+\bm{\nu}^{k})\right\|_2^2\nonumber\\
	&{\rm subject\;to} ~\mathbf{A}_t\mathbf{w}=\mathbf{u}
	\label{eq:admm3}\\
	&\!	\left[\!\begin{array}{l}
	\bm{\lambda}^{k+1}\\
	\bm{\nu}^{k+1}
	\end{array}\!\!\right]=\left[\!\begin{array}{l}
	\bm{\lambda}^{k}+(\tilde{\mathbf{w}}^{k+1}-\mathbf{w}^{k+1})\\
	\bm{\nu}^{k}+(\tilde{\mathbf{u}}^{k+1}-\mathbf{u}^{k+1})
	\end{array}\!\!\right]\label{eq:admm4}
	\end{align}
\end{subequations}
where all the dual variables have been scaled by the factor $\rho>0$ \cite{Boyd10}.

Interestingly enough, the solutions of \eqref{eq:admm1}-\eqref{eq:admm3} can be provided in closed form, as we elaborate in the following two propositions, whose proofs are deferred to the Appendix.
 \begin{proposition}\label{prop:16ab}
 	The solutions of \eqref{eq:admm1} and \eqref{eq:admm2} are respectively
 	\begin{subequations}\label{eq:admmc}
 		\begin{align}
 		\tilde{\mathbf{w}}^{k+1}&:=\frac{\rho}{1+\rho}\big(\mathbf{w}^k-\bm{\lambda}^k\big)\label{eq:admm11}\\
 		\!\tilde{\mathbf{u}}^{k+1}&:=\mathbf{c}_t\!+\mathcal{S}_{\nicefrac{1}{2\rho}}\big(\Re(\mathbf{u}^k\!-\bm{\nu}^k)\!-\mathbf{c}_t\big)\!+i \Im(\mathbf{u}^k-\bm{\nu}^k)\label{eq:admm21}
 		\end{align}
 	\end{subequations}
where the shrinkage operator $\mathcal{S}_\tau(\mathbf{x}):\mathbb{R}^N\times \mathbb{R}_+\to\mathbb{R}^N$ is $\mathcal{S}_\tau(\mathbf{x}):={\rm sign}(\mathbf{x})\odot\max(|\mathbf{x}|-\tau\mathbf{1},0)$, with $\odot$ and $|\cdot|$ denoting the entrywise multiplication and absolute operators, respectively, and ${\rm sign}(x):=\Big\{\!\!\begin{array}{ll}\nicefrac{x}{|x|},&x\ne 0\\
 	0,&x=0
 	\end{array} \Big.\!$ provides an entrywise definition of operator ${\rm sign}(\mathbf{x})$.
 \end{proposition}


The constrained minimization of \eqref{eq:admm3} essentially projects the pair  $(\tilde{\mathbf{w}}^{k+1}+\bm{\lambda}^{k},\tilde{\mathbf{u}}^{k+1}+\bm{\nu}^{k})$
onto the convex set specified by the linear equality constraint, namely $\{(\mathbf{w},\mathbf{u}):\mathbf{A}_t\mathbf{w}=\mathbf{u}\}$. Its solution is derived in a simple closed form next.
\begin{proposition}\label{pr:proj}
	Given $\mathbf{b}\in\mathbb{C}^N$ and $\mathbf{d}\in\mathbb{C}^M$, the solution of 
	\begin{align*}
			\underset{\mathbf{w}\in\mathbb{C}^N,\,\mathbf{u}\in\mathbb{C}^M}{\minimize}~&~\frac{1}{2}\left\|\mathbf{w}-\mathbf{b}\right\|_2^2+\frac{1}{2}\left\|\mathbf{u}-\mathbf{d}\right\|_2^2\\
	{\rm subject~to\,\,}~&~\mathbf{A}\mathbf{w}=\mathbf{u}
	\end{align*}
	is given as 
	\begin{subequations}\label{eq:projs}
		\begin{align}
		{\mathbf{w}^\ast}&:=\left(\mathbf{I}+\mathbf{A}^\ccalH\mathbf{A}\right)^{-1}\left(\mathbf{b}+\mathbf{A}^\ccalH\mathbf{d}\right)\label{eq:projs1}\\
		{\mathbf{u}^\ast}&:=\mathbf{A}{\mathbf{w}^\ast}\label{eq:projs2}.
		\end{align}
	\end{subequations}
\end{proposition}

Using Proposition \ref{pr:proj}, the minimizer of \eqref{eq:admm3} is found as
	\begin{subequations}\label{eq:inve}
		\begin{align}
		\mathbf{w}^{k+1}&:=(\mathbf{I}+\!\mathbf{A}_t^\ccalH\mathbf{A}_t)^{\!-1} \!\left[	\tilde{\mathbf{w}}^{k+1}\!+\bm{\lambda}^k \!+\mathbf{A}^\ccalH( \tilde{\mathbf{u}}^{k+1}\!+\bm{\nu}^k)\right]\label{eq:inve1}\\
		\mathbf{u}^{k+1}&:=\mathbf{A}\mathbf{w}^{k+1}.\label{eq:inve2}
		\end{align}
	\end{subequations}

The four updates in \eqref{eq:admm}
 are computationally simple except for the matrix inversion of \eqref{eq:inve1}, which nevertheless can be cached once computed during the first iteration. 
In addition, variables $\tilde{\mathbf{u}}^0$, $\bm{\lambda}^0$, and $\bm{\nu}^0$
of ADMM can be initialized to zero. Finally, the solution of \eqref{eq:midd} can be obtained as 
\begin{equation}\label{eq:dlav}
\mathbf{v}_{t+1}:=\mathbf{v}_t+\mathbf{w}^{\ast}
\end{equation}
 where $\mathbf{w}^\ast$ is the converged $\mathbf{w}$-iterate of the ADMM iterations in \eqref{eq:admmc},  \eqref{eq:inve}, and \eqref{eq:admm4}.

 The novel deterministic LAV solver based on ADMM is summarized in Table \ref{alg:dlav}, in which the inner loop consisting of Steps $3$-$8$ can be replaced with off-the-shelf solvers to solve \eqref{eq:midd} for $\mathbf{v}_{t+1}$. 
 
As far as performance is concerned, if $h$ is $L$-Lipschitz and $\nabla c$ is $\beta$-Lipschitz, then taking a constant stepsize $\mu\le {1}/(L\beta)$ in \eqref{eq:upda} guarantees that \cite{2016plconvergence}:

 i) the proposed solver in Table \ref{alg:dlav} is a descent method; and, 

 ii) the iterate sequence $\{\mathbf{v}_t\}$ converges to a stationary point of the LAV objective in \eqref{eq:lav}.

 The computational burden of the ADMM based deterministic solver is dominated by the projection step of \eqref{eq:inve}, which incurs per-iteration computational complexity on the order of $\mathcal{O}(MN^2)$. 
This complexity can be afforded in small- or medium-size PSSE tasks, but may not be efficient enough for nowadays increasingly interconnected power networks. This motivates our stochastic alternative of the ensuing section that relies on very inexpensive iterations.

 \begin{table}[t]
\renewcommand{\arraystretch}{1.2}
\caption{Deterministic LAV Solver Using ADMM}\label{alg:dlav}
\small
\vspace{-0.3em}
\begin{tabular}{|p{0.93\linewidth}|}
\hline
\vspace{-.3em}
1: Input data $\{(z_m,\mathbf{H}_m)\}_{m=1}^M$, stepsizes $\mu$, $\rho>0$, and initialization $\mathbf{v}_0=\mathbf{1}$.\\
2: \textbf{For} $t=0,1,\ldots,$ \textbf{do}\\
3:\hspace*{.6em} Initialize $\mathbf{w}^{0}$, $\mathbf{u}^0$, $\bm{\lambda}^0$, and $\bm{\nu}^0$ to zero.\\
4:\hspace*{.6em} Evaluate $\mathbf{A}_t$ and $\mathbf{c}_t$ as in \eqref{eq:dc}.\\
5:\hspace*{.6em} \textbf{For} $k=0,1,\ldots,$ \textbf{do}\\
6:\hspace*{1.2em} Update $(\tilde{\mathbf{w}}^{k+1},\tilde{\mathbf{u}}^{k+1})$, $(\mathbf{w}^{k+1},\mathbf{u}^{k+1})$, and $(\bm{\mu}^{k+1},\bm{\nu}^{k+1})$ using \eqref{eq:admmc},  \eqref{eq:inve}, and \eqref{eq:admm4}, respectively.\\
7:\hspace*{.6em} \textbf{End for}\\
8:\hspace*{.6em} Update $\mathbf{v}_{t+1}$ via \eqref{eq:dlav}.\\
9: \textbf{End for}\vspace{.5em}\\
\hline
\end{tabular}
\vspace{-0.5em}
\color{black}
\end{table}

\section{Stochastic Prox-linear LAV Solver}\label{sec:slav}

Finding the minimizer of \eqref{eq:midd} exactly per iteration of the deterministic LAV solver
may be computationally expensive, and can be intractable when the network size grows very large. 
Considering the wide applicability of LAV estimation as well as the increasing interconnection of microgrids, scalable online and stochastic approaches become of substantial interest. In this section, a stochastic linear proximal algorithm of 
\cite{duchi2017stochastic} is adapted to our PSSE task, and enables the prox-linear method to efficiently solve the LAV estimation problem at scale. Advantages of the stochastic approaches over their deterministic counterparts include oftentimes simple closed-form updates as well as faster convergence to yield an (approximately) optimal solution. Leveraging the sparsity structure of the measurement matrices and judiciously grouping measurements into small mini-batches can considerably speedup implementation of the stochastic solver. 

\subsection{Stochastic LAV solver}
Instead of dealing with the quadratic subproblems in \eqref{eq:midd}, each iteration of the stochastic LAV solver samples a datum $m_t\in \{1,2,\ldots,M\}$ uniformly at random from the total $M$ of measurements, and substitutes functions $(h,c)$ by $(h_{m_t},c_{m_t})$ associated with the sampled datum in the local linearization of \eqref{eq:llin}, hence also in \eqref{eq:midd}, yielding
\begin{equation}
\label{eq:supd}
\mathbf{v}_{t+1}\!:=\!	\underset{\mathbf{v}\in\mathbb{C}^N}{\arg\min} \left\{\left|\Re(\mathbf{a}_{m_t}^\ccalH(\mathbf{v}-\mathbf{v}_t))-\!c_{m_t}\right|+\frac{1}{2\mu_t}\left\|\mathbf{v}-\mathbf{v}_{t}\right\|_2^2 \right\}
\end{equation}
where the coefficients are given by
 \begin{subequations}\label{eq:sc}
 	\begin{align}
 	 \mathbf{a}_{m_t}&:=2\mathbf{H}_{m_t}\mathbf{v}_t\label{eq:sca}\\
 	 c_{m_t}&:=z_{m_t}-\mathbf{v}_t^\ccalH\mathbf{H}_{m_t}\mathbf{v}_t\label{eq:scc}.
 	\end{align}
 \end{subequations}

Different from iteratively seeking solutions of \eqref{eq:midd} based on ADMM iterations, the minimization of \eqref{eq:supd} admits a simple closed-form minimizer presented in the next result, which is proved in the Appendix.
\begin{proposition}\label{pr:supd}
	Given $\mathbf{a}\in\mathbb{C}^N$ and $c\in\mathbb{R}$, the solution of 
	\begin{equation}\label{eq:thre}
		\underset{\mathbf{w}\in\mathbb{C}^N}{\minimize}~~\left|\Re(\mathbf{a}^\ccalH\mathbf{w})-c\right|+\frac{1}{2\tau}\|\mathbf{w}\|_2^2
	\end{equation}
	is given by $\hat{\mathbf{w}}:={\rm proj}_{\tau}(c/\|\mathbf{a}\|_2^2)\cdot\mathbf{a}$, where the projection operator ${\rm proj}_{\tau}(x):\mathbb{R}\times\mathbb{R}_+\to \mathbb{R}$ returns the real number in interval $[-\tau,\tau]$ closest to any given $x\in\mathbb{R}$. 
\end{proposition}

Based on Proposition \ref{pr:supd}, the solution of \eqref{eq:supd} is given by
\begin{equation}
\label{eq:slav}
\mathbf{v}_{t+1}:=\mathbf{v}_t+{\rm proj}_{\mu_t}\!(c_{m_t}/\|\mathbf{a}_{m_t}\|_2^2)\cdot\mathbf{a}_{m_t}.
\end{equation}
Intuitively, measurements with a relatively small absolute residual, namely $|c_{m_t}|\le \|\mathbf{a}_{m_t}\|_2^2$, are deemed `nominal,' and $\mathbf{v}_t$ is updated with a step of $c_{m_t}/\|\mathbf{a}_{m_t}\|_2^2$ along the current direction of $\mathbf{a}_{m_t}$. On the other hand, the measurements of larger absolute residuals are likely to be outliers, so $\mathbf{v}_t$ is updated along its direction $\mathbf{a}_{m_t}$ by only a step of $\tau_t$ as opposed to $c_{m_t}/\|\mathbf{a}_{m_t}\|_2^2$.


 \begin{table}[t]
 	\renewcommand{\arraystretch}{1.2}
 	\caption{Stochastic LAV Solver}\label{alg:slav}
 	\small
 	\vspace{-0.3em}
 	\begin{tabular}{|p{0.93\linewidth}|}
 		\hline
 		\vspace{-.3em}
 		1: Input data $\{(z_m,\mathbf{H}_m)\}_{m=1}^M$, stepsize $\{\mu_t>0\}$, and initialization $\mathbf{v}_0=\mathbf{1}$.\\
 		2: \textbf{For} $t=0,1,\ldots $ \textbf{do}\\
 		3:\hspace*{1em}Draw $m_t$ uniformly at random from $\{1,2,\ldots,M\}$, or, by cycling through $\{1,2,\ldots,M\}$.\\
 	4:\hspace*{1em}Evaluate  $\mathbf{a}_{m_t}$ and $c_{m_t}$ as in \eqref{eq:sc}.\\
 		5:\hspace*{1em} Update $\mathbf{v}_{t+1}$ using \eqref{eq:slav}.\\
 		6: \textbf{End for} 	\vspace{.5em}\\
 		\hline
 	\end{tabular}
 	\vspace{-0.5em}
 	\color{black}
 \end{table}

The proposed stochastic prox-linear LAV solver is listed in Table \ref{alg:slav}. For convergence, a diminishing stepsize sequence $\{\mu_t\}$ is required. Specifically, we consider stepsizes that are square summable but not summable; that is,
\begin{equation}
\sum_{t=0}^\infty\mu_t=\infty,\quad {\rm and} \quad \sum_{t=0}^\infty\mu_t^2<\infty.
\end{equation}
For instance, one can choose $\mu_t=\alpha t^{-\beta}$ with appropriately selected constants $\alpha>0$ and $\beta\in (0.5,1]$. Then the sequence $\{\mathbf{v}_{t}\}$ converges to a stationary point of the LAV objective in $\eqref{eq:lav}$ almost surely \cite[Thm. 1]{duchi2017stochastic}. 

In terms of computational complexity, it can be verified that each $\mathbf{H}_{m}$ matrix corresponding to a square voltage magnitude or (re)active power flow measurement [cf. \eqref{eq:squa} and \eqref{eq:flow}] has exactly one or three nonzero entries, respectively. As such, if the available measurements include only these two types of measurements,  
	evaluating $(\mathbf{a}_{m_t},c_{m_t})$ as well as updating $\mathbf{v}_t$ per stochastic iteration requires just a small number ($\le\!10$) of scalar multiplications and additions, therefore incurring per-iteration complexity of $\mathcal{O}(1)$, which is \emph{independent of the network size $N$}.
	This complexity evidently scales favorably to very large interconnected power networks. It is also worth highlighting that only one or two entries of $\mathbf{v}_t$ are updated depending on whether a voltage or power flow measurement is processed at each iteration.
	On the other hand, even if the power injections are measured too, 
	the number of scalar operations per iteration increases to the order of the number of neighboring buses, which still remains much smaller than $N$ in most real-world networks.

   \begin{remark}
   PMU measurements can be easily accounted for in \eqref{eq:lav}. The developed prox-linear LAV schemes apply without any algorithmic modification.  
   \end{remark}

\begin{figure*}
	\begin{minipage}[b]{0.5\linewidth}
		\centering
				\includegraphics[width = 5.6cm, height=5cm]{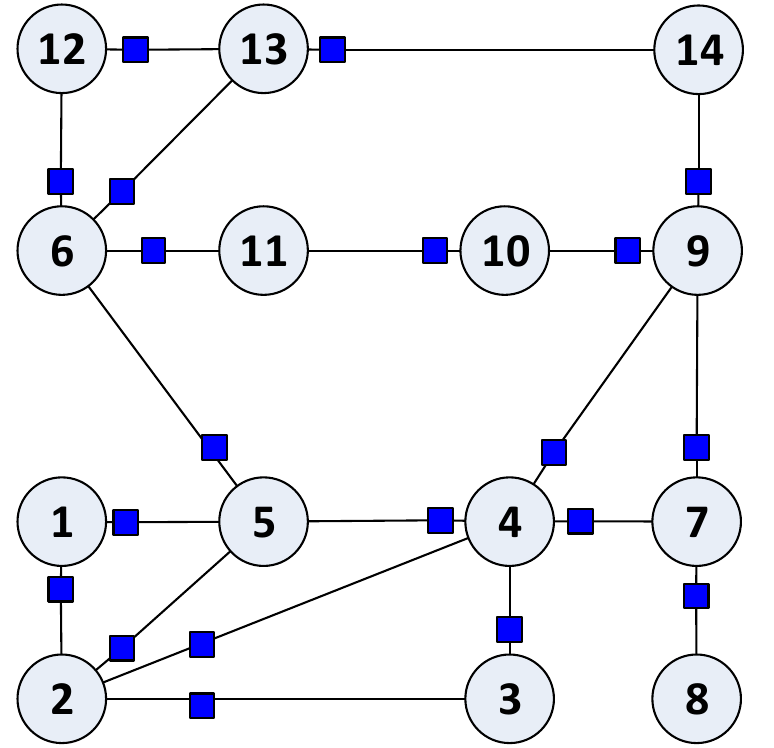} 
			\vspace{.28cm}
				\caption{The IEEE $14$-bus benchmark configuration.}
							\label{fig:ieee14}
			\end{minipage}
			\begin{minipage}[b]{.5\linewidth}
				\renewcommand{\arraystretch}{1.6}
				\centering
						\begin{tabular}{ c| | l }
							\hline
														Groups & Power flow measurements (Bus index) \\ \hline
							\hline
							$1$& $(1,2)$, $(3,4)$, $(5,6)$, $(7,9)$ \\ \hline
							$2$&$(1,5)$, $(2,4)$, $(6,11)$, $(12,13)$\\ \hline
							$3$ & $(2,3)$, $(4,5)$, $(6,12)$, $(9,14)$ \\ \hline
							$4$&$(2,5)$, $(4,7)$, $(6,13)$, $(10,11)$ \\ \hline
							$5$ &$(4,9)$, $(7,8)$, $(9,10)$, $(13,14)$ \\    \hline
						\end{tabular}
						\vspace{.5cm}
					\captionof{table}{Mini-batching power flow measurements.}
											\label{tab:group}
			\end{minipage}

\end{figure*}



\subsection{Accelerated implementation using mini-batches}
\label{sec:acce}
Although it involves only simple closed-form updates, the stochastic solver in Table \ref{alg:slav} may require a large number of iterations to converge for high-dimensional power networks. Stochastic approaches based on mini-batches of measurements have been recently popular in large-scale machine learning tasks, because they offer a means of accelerating the stochastic algorithms. Yet, the naive way of designing mini-batches by grouping measurements randomly would yield a sequence of quadratic programs as in \eqref{eq:midd} of the deterministic LAV solver, which does not have closed-form solutions due to the $\ell_1$ term. The novelty here is to fully exploit the sparsity of $\mathbf{H}_m$ matrices to group measurements into mini-batches in a way  that closed-form solutions of the resulting quadratic programs are possible. 
 
Suppose that active and reactive power flows over all lines and the square voltage magnitude at all buses are measured. 
Since $\mathbf{H}_n^V$ in \eqref{eq:squa} has exactly one nonzero entry at the $(n,n)$-th position, the corresponding updating vector $\mathbf{a}_n$ in \eqref{eq:sca} has (at most) one nonzero entry at the $n$-th position. Updating $\mathbf{v}_t$ using \eqref{eq:slav} modifies the $n$-th position only. 
Hence, refining $\mathbf{v}$ using the $N$ voltage measurements sequentially in $N$ stochastic iterations is equivalent to updating $\mathbf{v}$ using all $N$ measurements simultaneously in a single iteration.
For power flows, every $\mathbf{H}_m$ has three nonzero entries in two rows. For instance, $\mathbf{H}^P_{nn'}$ has nonzero entries indexed by $(n,n')$, $(n,n)$, and $(n',n)$; and so does $\mathbf{H}_{nn'}^Q$. Processing the active or reactive power flow measurement over line $(n,n')$ amounts to updating the $n$-th and $n'$-th entries of $\mathbf{v}_t$. Hence, so long as each of a mini-batch of measurements does not share common indices with the remaining ones, processing a mini-batch of such measurements one after the other boils down to processing all measurements simultaneously in one iteration. 

 For illustration, consider the IEEE $14$-bus test system depicted in Fig. \ref{fig:ieee14} \cite{PSTCA}. Consider a total of $54$ measurements, which include $14$
  square voltage magnitudes, as well as $20$ `from-end' active and reactive power flows each.
 All voltage magnitudes can be grouped as a single mini-batch, or into several mini-batches by any means. One way of mini-batching each type of power flow measurements is suggested in Table \ref{tab:group}, where $20$ active (reactive) power flows yield $5$ mini-batches of equal size. It can be easily verified that any two measurements within a group (mini-batch) are measured over two lines of non-overlapping indices. 
 
Let the entire measurements be divided into $B$ mini-batches denoted by $\{\mathcal{B}_b\}_{b=1}^{B}\subseteq\{1,2,\ldots,M\}$. If a mini-batch $\mathcal{B}_{b_t}$ of measurements is drawn uniformly at random from $\{\mathcal{B}_{b}\}_{b=1}^{B}$ at iteration $t$, the accelerated implementation by means of mini-batching, updates the state estimate according to [cf. \eqref{eq:slav}]
\begin{equation}
\label{eq:alav}
\mathbf{v}_{t+1}:=\mathbf{v}_t+\!\sum_{m\in\mathcal{B}_{b_t}}\!{\rm proj}_{\mu_t}(c_{m}/\|\mathbf{a}_{m}\|_2^2)\cdot\mathbf{a}_m
\end{equation}
   which is in sharp contrast to that of using ADMM iterations to deal with the quadratic subproblems \eqref{eq:upda} in the
   deterministic LAV solver.

\section{Numerical Tests}\label{sec:test}
The proposed linear proximal LAV solvers were numerically tested in this section. Three power network benchmarks including the IEEE $14$-, $118$-bus, and the PEGASE $9,241$-bus systems were simulated, following the MATLAB-based toolbox MATPOWER~\cite{PSTCA, MATPOWER}. 

The linear programming (LP) and the iteratively reweighted least-squares (IRLS) based LAV estimators \cite{1982baddata}, \cite{1991fast}, \cite{jabr2004}, along with the `workhorse' Gauss-Newton iterations for the WLS-based PSSE \cite{AburExpositoBook} were adopted as baselines. It is worth mentioning that the LP-based implementation can be regarded as a special case of the deterministic prox-linear algorithm with a constant step size of $\infty$. To see this, per iteration, the LP-based scheme \cite{1991fast} solves the minimization problem in \eqref{eq:upda} but \emph{without} the augmented majorization term $\frac{1}{2\mu_t}\|\mathbf{v}-\mathbf{v}_t\|_2^2$, or equivalently with $\mu_t=\infty$.
	To be specific, the linear program was formulated over $2N-1$ real variables consisting of the real and imaginary parts of the unknown voltage phasor vector, after excluding the imaginary part of the reference bus which was set $0$. Per iteration, the resultant linear program was solved by calling for the convex optimization package CVX \cite{2008cvx} together with its embedded interior-point solver SeDuMi \cite{sedumi}. Given that there is no parameter in the LP-based LAV estimator, although the time performance may vary if different toolboxes are used for solving the resultant linear programs, its convergence behavior in terms of the number of iterations is \emph{independent} of the toolbox used. 
	Furthermore,
 the Gauss-Newton iterations were implemented by calling for the embedded state estimation function `doSE.m' in MATPOWER.
  
 Regarding the initialization, when all squared voltage magnitudes are measured, the initial point is taken to be the voltage magnitude vector, unless otherwise specified. 
Each simulated scheme stops either when a maximum number $100$ of iterations are reached, or when the normalized distance between two consecutive estimates becomes smaller than $10^{-10}$, namely $\|\mathbf{v}_t-\mathbf{v}_{t-1}\|_2/\sqrt{N}\le 10^{-10}$. 
In order to fix the phase ambiguity, the phase generated at the reference bus was set to $0$ in all tests.
For numerical stability, and to eliminate the effect of certain leverage measurements \cite{2014abur}, the developed solvers were implemented using the normalized data, namely $\{(\frac{z_m}{\|\mathbf{H}_m\|_2},\frac{\mathbf{H}_m}{\|\mathbf{H}_m\|_2})\}_{m=1}^M $. Although this work focused on fast and scalable implementations of LAV estimators, certain enhanced solvers that possess similar compositional structure as in LAV \eqref{eq:lav} 
can also benefit from the developed composite optimization algorithms. Those include e.g., the (robustified) Schweppe-Huber generalized M-estimator \cite{mili1996robust}.

\subsection{Noiseless case}

The first experiment simulates the noiseless data to evaluate the convergence and runtime of the novel algorithms relative to the WLS-based Gauss-Newton iterations, as well as the LP- and IRLS-based LAV solvers on the IEEE $14$-bus test system. The default voltage profile was employed. 
Measurements including all (`sending-end') active and reactive power flows, as well as all squared voltage magnitudes were obtained from MATPOWER \cite{MATPOWER}. 
The ADMM-based deterministic prox-linear solver in Table \ref{alg:dlav} was implemented with stepsize $\mu=200$, where each quadratic subproblem was solved using a maximum of $150$ ADMM iterations with stepsize $\rho=100$. 
It is worth mentioning that the deterministic prox-linear solver can be also implemented using standard convex programming approaches (by solving subproblem \eqref{eq:upda} directly). It typically converges in a few (less than $10$) iterations yet at a higher computational complexity. 
The stochastic algorithm in Table \ref{alg:slav} used the diminishing stepsize $ 1/ t^{0.8}$.   
The accelerated scheme in \eqref{eq:alav} was implemented with stepsize $0.8$ using a total of $11$ mini-batches: $1$ for all voltage magnitudes, and $5$ of equal size for (sending-end) active and reactive power flows, each grouped as in Table \ref{tab:group}. The normalized root mean-square error (RMSE) $\|{\mathbf{v}}_t-\mathbf{v}\|_2/\|\mathbf{v}\|_2$ was evaluated at every Gauss-Newton iteration, per linear program, and every $M$ stochastic iterations of the stochastic and accelerated schemes, where $\mathbf{v}$ is the true voltage profile, and ${\mathbf{v}}_t$ denotes the estimate obtained at the $t$ iteration.

\begin{figure}[t]
		\vskip 0.1in
		\centering
		\includegraphics[scale = 0.58]{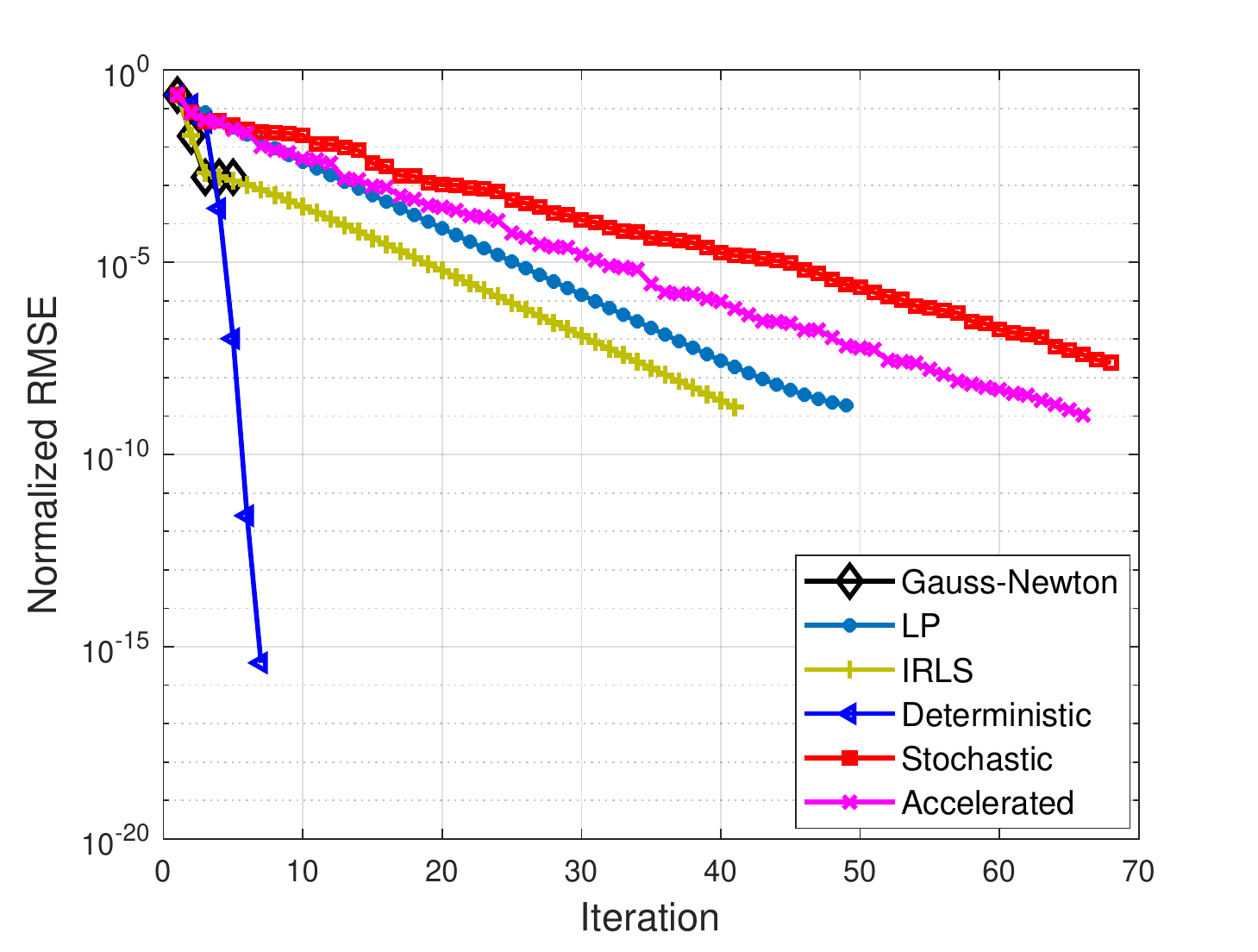} 
		\caption{Convergence performance for the IEEE $14$-bus system.
		}
		\label{fig:1st}
		\vskip -.05in
	\end{figure}

Figure \ref{fig:1st} compares the normalized RMSE for the LP-, and IRLS-based, deterministic, stochastic, and accelerated LAV solvers with that of the WLS-based Gauss-Newton iterations, whose corresponding runtime and number of iterations to reach the stopping criterion are tabulated in Table \ref{tab:comparison}.
Evidently, the deterministic scheme is the fastest in terms of both the number of iterations and runtime, and converges to a point of machine accuracy (i.e., $10^{-16}$) in $8$ iterations. The IRLS is also fast, but similar to the WLS-based Gauss-Newton iterations, it requires inverting a matrix per iteration which may discourage its use in large power systems. 
Even though the time of solving each LP may vary across toolboxes, convergence of the LP-based scheme in terms of the number of iterations will be the same. Evidently, solving a LP of $2M$ constraints and $2N-1$ real variables is computationally more cumbersome and slower than performing $M$ (accelerated) stochastic LAV iterations, hence justifying the fast convergence rate of the proposed LAV solvers.

The Gauss-Newton method terminated after six iterations, but at a sub-optimal point of normalized RMSE $10^{-3}$ or so. The accelerated implementation is comparable with the stochastic LAV solver, and yields an accurate solution with RMSE $=4.28\times 10^{-8}$ in time also comparable to the Gauss-Newton iterations.
The proposed 
	 LAV solvers are much faster than the LP-based implementation.

	\begin{table}[t]	 
		\begin{center}
	\captionof{table}{Comparisons of Different State Estimators}\label{tab:comparison}\vspace{.5em}
			\begin{tabular}{ c | c | c }
				\hline
				Algorithm & Number of iterations & Computation time (s) \\ \hline
				Linear Program~\cite{1982baddata}& $49$ & $15.167$ \\ \hline
			IRLS~\cite{jabr2004}& $42$ & $0.074$ \\ \hline
				Gauss-Newton~\cite{AburExpositoBook}&$5$ & $0.071$\\ \hline
				Deterministic LAV & $6$ & $ 0.062$ \\ \hline
				Stochastic LAV&$68$ &$ 0.110$ \\ \hline
				Accelerated &$66$ &$0.103$ \\ \hline
			\end{tabular}
			\vspace{.5em}
		\end{center}
	\end{table}

	\subsection{Presence of outliers}

\begin{figure}[t]
			\vskip -0.0in
			\centering
		\includegraphics[scale = 0.58]{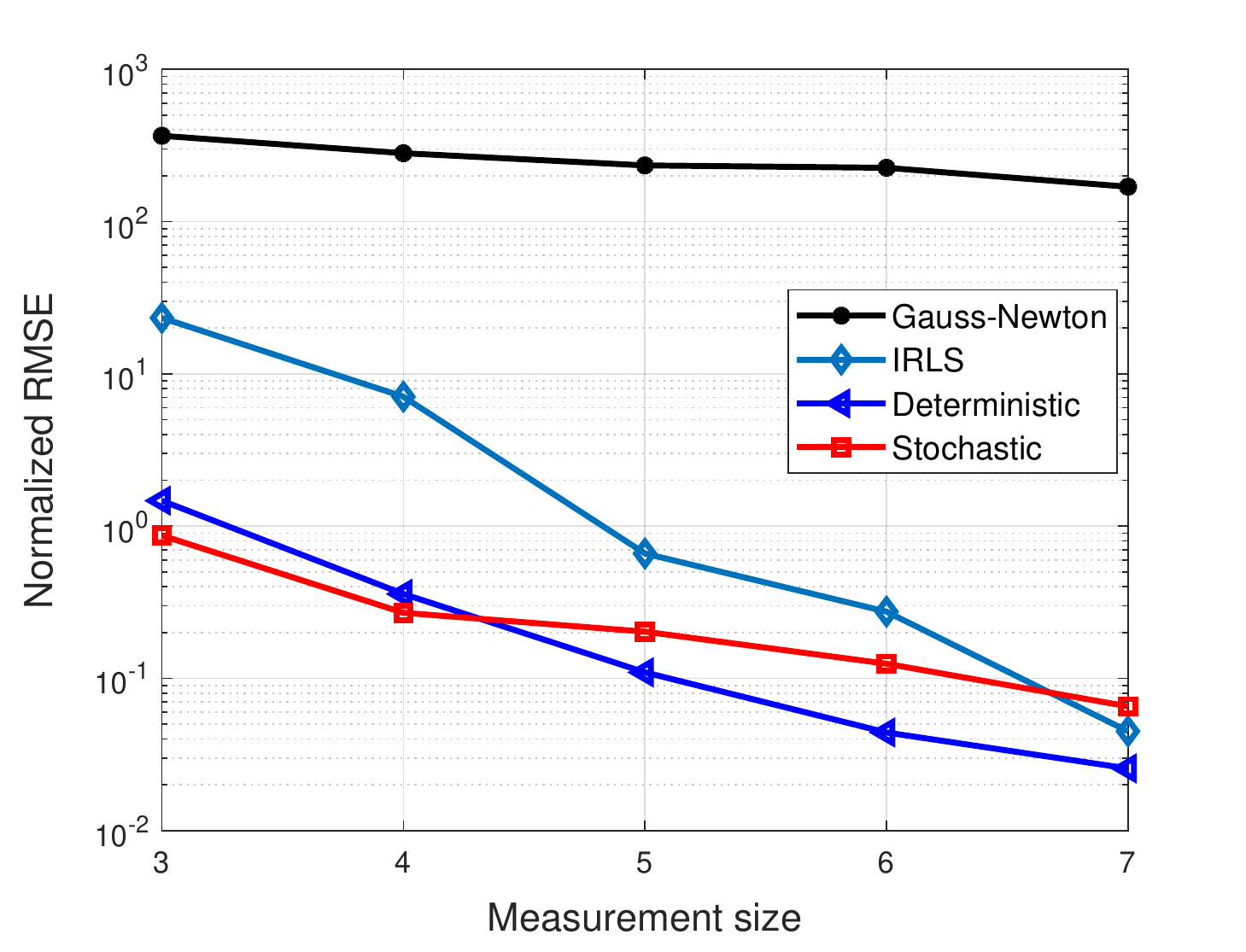} 
			\caption{Robustness to outliers for the IEEE $118$-bus system.
			}
			\label{fig:3rd}
			\vskip .05in
		\end{figure}

	The second experiment was set to assess the robustness of the novel
	deterministic and stochastic solvers to measurements with outliers using the IEEE $118$-bus benchmark network \cite{PSTCA}, while
	the IRLS-based LAV implementation and the WLS-based Gauss-Newton iterations were simulated as baselines.
	 The actual voltage magnitude (in p.u.) and angle (in radians) of each bus were uniformly distributed over $[0.9,\,1.1]$, and over $[-0.1\pi,\,0.1\pi]$. To assess the PSSE performance versus the measurement size, an additional type of measurements was included in a deterministic manner, as described next.
	All seven types of SCADA measurements were first enumerated as
	$\{|V_n|^2,\,P_{nk}^f,\,Q_{nk}^f,\,P_{n},\,Q_{n},\,P_{nk}^t,\,Q_{nk}^t\} $.
	Each $x$-axis value in Fig. \ref{fig:3rd} signifies the number of ordered types of measurements used in the experiment to yield the corresponding normalized RMSEs, which are obtained by averaging over $100$ independent realizations. For example, $5$ implies that the first $5$ types of data (i.e., $|V_n|^2,P_{nk}^f,Q_{nk}^f,P_{n},Q_{n}$ over all buses and lines) were measured.
	Additive noise was independently generated from Gaussian distributions having zero-mean and standard deviation $0.004$, $0.008$, and $0.01$ p.u. for the voltage magnitude, line flow, and power injection measurements, respectively \cite{1991fast}. Ten percent of the measured data were corrupted according to model M1, chosen randomly from line flows and bus injections. The outlying data $\{\xi_m\}$ were drawn independently from a Laplacian distribution with zero-mean and standard deviation $30$. The subproblems \eqref{eq:upda} with $\mu=100$ of the deterministic scheme were solved using a maximum of $200$ ADMM iterations with stepsize $\rho=100$, while the stochastic one was implemented with a diminishing stepsize $\mu_t=10/t^{0.9}$. It is evident from Fig. \ref{fig:3rd} that our prox-linear LAV schemes are  resilient to outlying measurements under corruption model M1, yielding improved performance relative to  the IRLS estimator. Furthermore, IRLS works well when the number of measurements grows large. Finally, the ADMM-based prox-linear estimator requires the least number of iterations for convergence. It is followed by the IRLS estimator, and then by the stochastic prox-linear estimator.

			\begin{table}[hbt]	 
						\vskip .2in
				\begin{center}
							{\color{black}	
					\captionof{table}{Comparisons of the Gauss-Newton and Stochastic LAV Estimators}\label{tab:exp3}\vspace{.5em}
					\begin{tabular}{ c | c | c }
						\hline
						Algorithm & Number of iterations & Computation time \\ \hline
						Gauss-Newton~\cite{AburExpositoBook}&$10$ (maximum) & $\ge 3$hs\\ \hline
						Stochastic LAV&$22$ &$ 2,595$s \\ \hline
					\end{tabular}
				}
					\vspace{.5em}
				\end{center}
			\end{table}

		The third experiment tests the scalability and efficacy of the stochastic iterations on a larger power network of $9,241$ buses available in MATPOWER \cite{MATPOWER}. The true voltage magnitude of each bus was uniformly distributed over $[0.95,\,1.05]$, and its angle  over $[-0.05\pi,\,0.05\pi]$. The maximum number of iterations for the Gauss-Newton method was set $10$. 
		All seven types of SCADA data were measured with additive noise described in the last experiment, 
		$5\%$ of which were compromised under model M2. The corrupted data $\xi_m:=\tilde{\mathbf{v}}^\ccalH\mathbf{H}_m\tilde{\mathbf{v}}$ relying on the individual $\mathbf{H}_m$ were generated using $\tilde{\mathbf{v}}\in\mathbb{R}^N$ from the standardized multivariate Gaussian distribution. 
In words, there were a total of $18,481$ variables to be estimated, 
 a total of $91,919$ measurements were obtained, $4,595$ of which were purposefully manipulated by adversaries. Initialized with the flat voltage profile point, the WLS-based Gauss-Newton iterations yielded an estimate of RMSE $0.9846$, whereas the stochastic scheme in Table \ref{alg:slav} with diminishing stepsize $\mu_t=100/t^{0.8}$ attained an RMSE of $0.0412$. 
The corresponding computational runtime of each scheme was reported in Table \ref{tab:exp3}.
 Evidently, the
 stochastic LAV implementation is several times faster than the WLS-based Gauss-Newton iterations.

 	\begin{remark}\label{rmk:comparison}
 		 Each Gauss-Newton iteration involves inverting a $(2N-1)\times (2N-1)$ matrix,  which incurs computational complexity $\mathcal{O}((2N-1)^3)$. It is clear when the system size $N$ grows large, say $N\ge 10,000$, this cubic complexity of Gauss-Newton iterations as well as the memory required may easily become prohibitive for a desktop computer. On the contrary, the per-iteration complexity of the proposed stochastic LAV scheme can be as low as $\mathcal{O}(1)$, which is clearly scalable, and well-tailored for PSSE tasks of large dimensions. It is thus intuitive that in large-scale power systems, the proposed stochastic iterations based LAV implementation is faster than the Gauss-Newton iterations. The advantage of adopting inexpensive stochastic iterations to handle large-scale optimization problems has been corroborated by the recent success of deep learning for visual recognition and speech translation too, where stochastic gradient based approaches (e.g., stochastic gradient descent) constitute the `workhorse' in training deep neural networks \cite{2015deep}. 
 	\end{remark}

\section{Conclusions}
\label{sec:conc}

Robust power system state estimation was pursued using contemporary tools of composite optimization. Building on recent algorithmic advances, two solvers were put forward to efficiently handle the LAV-based PSSE. Specifically, a deterministic LAV method was developed based on a linear proximal method, which yields a sequence of convex quadratic subproblems that can be efficiently solved using off-the-shelf solvers, or, through fast ADMM iterations. It converges as fast as Gauss-Newton iterations, amounting to solving only $8\sim10$ quadratic programs in general. 
Inspired by the sparse connectivity inherent to power networks, a highly scalable stochastic scheme that can afford simple closed-form updates was also devised. When only line flows and voltage magnitudes are measured, each stochastic iteration performs merely a few complex scalar operations, incurring per-iteration complexity $\mathcal{O}(1)$, regardless of the number of buses in the entire network. 
If, on the other hand, the power injections are included as well, this time complexity goes down to the order of the number of neighboring buses, which still remains much smaller than the network size in general. 
A mini-batching technique was suggested to further accelerate the stochastic iterations by means of leveraging the sparsity of measurement matrices. 
Numerical tests on a variety of benchmark networks of up to $9,241$ buses showcase the robustness and computational efficiency of the developed approaches relative to existing alternatives, particularly over large-size networks. 

Devising decentralized and parallel implementations of the novel approaches constitutes interesting future research directions. Since the LAV estimator may yield non-robust estimates in the presence of bad leverage points, and the measurement scaling may not be able to effectively identify and eliminate a certain type of leverage measurements, it is meaningful and promising to generalize the presented deterministic and stochastic proximal-linear based algorithmic tools to other robustness-enhanced estimators such as the Schweppe-Huber generalized M-estimator \cite{mili1996robust}, \cite{tps2018zm}, \cite{tps2018zmp}.
Coping with the $Y$- and $\Delta$-connection, as well as investigating the technical approaches in multiphase unbalanced distribution systems are practically relevant future research topics too.

\section*{Appendix}
\begin{IEEEproof}[Proof of Prop. \ref{prop:16ab}]
	It is easy to check that the solution of \eqref{eq:admm1} is given by \eqref{eq:admm11}, whose proof is thus omitted. 
Considering any $\mathbf{c}\in\mathbb{R}^N$ and $\mathbf{d}\in\mathbb{C}^N$, 
solving \eqref{eq:admm2} is equivalent to solving 
\begin{equation}
	{\mathbf{u}}^\ast:=\underset{{\mathbf{u}}\in\mathbb{C}^N}{\arg\min}~\lambda\!\left\|\Re({\mathbf{u}})-\mathbf{c}\right\|_1+\frac{1}{2}\left\|{\mathbf{u}}-\mathbf{d}\right\|_2^2	\label{eq:admm2app}.
\end{equation}
Upon defining $\mathbf{x}:=\mathbf{u}-\mathbf{c}$, problem \eqref{eq:admm2app} becomes
\begin{equation*}
	\min_{\mathbf{x}\in\mathbb{C}^N}\;\lambda\!\left\|\Re({\mathbf{x}})\right\|_1+\frac{1}{2}\!\left\|{\mathbf{x}}-(\mathbf{d}-\mathbf{c})\right\|_2^2	
\end{equation*}
or equivalently, 
\begin{equation}
	\min_{\mathbf{x}}\,\lambda\|\Re({\mathbf{x}})\|_1+\frac{1}{2}\|\Re(\mathbf{x})-\Re(\mathbf{d}-\mathbf{c}))\|_2^2+	\frac{1}{2}\|\Im(\mathbf{x})-\Im(\mathbf{d}-\mathbf{c}))\|_2^2.\label{eq:2app}
\end{equation}

Problem \eqref{eq:2app} can be decomposed into two subproblems that correspond to optimizing over the real- and imaginary parts of $\mathbf{x}=\Re(\mathbf{x})+j\Im(\mathbf{x}):=\mathbf{x}_r+j\mathbf{x}_i$; that is
\begin{equation}
\min_{\mathbf{x}_r\in\mathbb{R}^N}\;\lambda \|\mathbf{x}_r\|_1+\frac{1}{2}\|\mathbf{x}_r-\Re(\mathbf{d}-\mathbf{c})\|_2^2\label{eq:2appr}
\end{equation}
and 
\begin{equation}
\min_{\mathbf{x}_i\in\mathbb{R}^N}\;\frac{1}{2}\|\mathbf{x}_i-\Im(\mathbf{d}-\mathbf{c})\|_2^2\label{eq:2appi}.
\end{equation}

The optimal solutions of the convex programs in  \eqref{eq:2appr} and \eqref{eq:2appi} can be found as, see e.g. \cite{Boyd10}
\begin{align*}
\mathbf{x}_r^\ast&:=\mathcal{S}_{\lambda}\big(\Re(\mathbf{d}-\mathbf{c})\big)
=\mathcal{S}_{\lambda}\big(\Re(\mathbf{d})-\mathbf{c}\big)\\
\mathbf{x}_i^\ast&:=\Im(\mathbf{d}-\mathbf{c})=\Im(\mathbf{d})
\end{align*}
thus yielding the optimal solution of \eqref{eq:2app} as
$$\mathbf{x}^\ast:= \mathbf{x}_r^\ast+j\mathbf{x}_i^\ast=\mathcal{S}_{\lambda}(\Re(\mathbf{d})-\mathbf{c})+j \Im(\mathbf{d}).$$
Recalling $\mathbf{u}=\mathbf{x}+\mathbf{c}$, the optimal solution of \eqref{eq:admm2app} is 
\begin{equation}
\mathbf{u}^\ast=\mathbf{x}^\ast+\mathbf{c}=\mathbf{c}+\mathcal{S}_{\lambda}(\Re(\mathbf{d})-\mathbf{c})+j \Im(\mathbf{d})
\end{equation}
which completes the proof. 
\end{IEEEproof}

\begin{IEEEproof}[Proof of Prop. \ref{pr:proj}] Letting $\bm{\chi}$ denote the dual variable associated with the constraint $\mathbf{u}=\mathbf{A}\mathbf{w}$, the KKT optimality conditions are given by  \cite{Boyd10}
		\begin{align*}
		\mathbf{w}^\ast-\mathbf{b}+\mathbf{A}^\ccalH\bm{\chi}^\ast&=\mathbf{0}\\
		\mathbf{u}^\ast-\mathbf{d}-\bm{\chi}^\ast&=\mathbf{0}\\
		\mathbf{A}\mathbf{w}^\ast-\mathbf{u}^\ast&=\mathbf{0}
		\end{align*}
	or in the following compact form
	\begin{equation*}
	\left[\begin{array}{ccc}
	\mathbf{I}_N&\mathbf{0}&\mathbf{A}^\ccalH\\	
	\mathbf{0}&\mathbf{I}_M&-\mathbf{I}_M\\
	\mathbf{A}&-\mathbf{I}_M&\mathbf{0}
	\end{array}
	\right]\left[\!\begin{array}{c}
	\mathbf{w}^\ast\\
	\mathbf{u}^\ast\\
	\bm{\chi}^\ast
	\end{array}
	\!\right]
	=\left[\begin{array}{c}
	\mathbf{b}\\
	\mathbf{d}\\
	\mathbf{0}
		\end{array}
	\right].
	\end{equation*}
	
	Eliminating the dual variable via $\bm{\chi}^\ast=\mathbf{u}^\ast-\mathbf{d}$ from the KKT system, yields
	\begin{equation}\label{eq:kt}
		\left[\begin{array}{cc}
		\mathbf{I}_N&\mathbf{A}^\ccalH\\	
		\mathbf{A}&-\mathbf{I}_M
		\end{array}
		\right]\left[\!\begin{array}{c}
		\mathbf{w}^\ast\\
		\mathbf{u}^\ast
		\end{array}
		\!\right]
		=\left[\begin{array}{c}
		\mathbf{b}+\mathbf{A}^\ccalH	\mathbf{d}\\
		\mathbf{0}
		\end{array}
		\right].
	\end{equation}
	
	By further eliminating $\mathbf{d}^\ast$ and solving for $\mathbf{b}^\ast$, the solution to \eqref{eq:kt} and also to the minimization in \eqref{eq:projs} can be found in two steps as
	\begin{align*}
		{\mathbf{w}^\ast}&:=\left(\mathbf{I}+\mathbf{A}^\ccalH\mathbf{A}\right)^{-1}\left(\mathbf{b}+\mathbf{A}^\ccalH\mathbf{d}\right)\\
		{\mathbf{u}^\ast}&:=\mathbf{A}{\mathbf{w}^\ast}
	\end{align*}
	which completes the proof of the claim.
	\end{IEEEproof}

\begin{IEEEproof}[Proof of Prop. \ref{pr:supd}] 
	The optimality condition for \eqref{eq:thre} is
	\begin{equation*}
	\mathbf{0}\in \partial \!\left(\left|\Re(\mathbf{a}^\ccalH\mathbf{w})-c\right|\right)+\frac{1}{\tau}\mathbf{w}\Longleftrightarrow \mathbf{0}\in\partial\! \left|\Re(\mathbf{a}^\ccalH\mathbf{w})-c\right|\cdot \mathbf{a}+\frac{1}{\tau}\mathbf{w}
	\end{equation*}
	or equivalently, $$\mathbf{0}\in \partial\! \left|\Re(\mathbf{a}^\ccalH\mathbf{w})-c\right|\cdot (\tau\mathbf{a})+\mathbf{w}, $$ where $\partial$ denotes the subdifferential. Let us first examine the case where $\Re(\mathbf{a}^\ccalH\mathbf{w})-c\ne 0$. We thus have $\partial|\Re(\mathbf{a}^\ccalH\mathbf{w})-c|={\rm sign}(\Re(\mathbf{a}^\ccalH\mathbf{w})-c)$, which yields the optimum
	\begin{equation*}
	\mathbf{w}^\ast=-\tau {\rm sign}(\Re(\mathbf{a}^\ccalH\mathbf{w})-c)\cdot\mathbf{a}.
	\end{equation*} 
Note that if ${c}/{\|\mathbf{a}\|_2^2}\ge \tau$, or
$\Re(\mathbf{a}^\ccalH\mathbf{w}^\ast)-c=-\tau\|\mathbf{a}\|_2^2\,{\rm sign}(\mathbf{a}^\ccalH\mathbf{w}^\ast-c)-c<0$, then $\mathbf{w}^\ast=\tau\mathbf{a}$. Equivalently, if ${c}/{\|\mathbf{a}\|_2^2}\le- \tau$, or
$\Re(\mathbf{a}^\ccalH\mathbf{w}^\ast)-c=-\tau\|\mathbf{a}\|_2^2\,{\rm sign}(\mathbf{a}^\ccalH\mathbf{w}^\ast-c)-c>0$, then $\mathbf{w}^\ast=-\tau\mathbf{a}$.

If $\Re(\mathbf{a}^\ccalH\mathbf{w})-c=0$, the subdifferential of the absolute-value operator belongs to the interval $[-1,1]$; hence, the optimality condition becomes
\begin{equation*}
\mathbf{0}\in -[-1,\;1]\cdot (\tau \mathbf{a})+\mathbf{w}\Longleftrightarrow \mathbf{w}\in [-\tau,\;\tau]\,\mathbf{a}.
\end{equation*}

Upon letting ${\rm proj}_\tau(x)$ denote the projection of a real number $x$ onto the interval $[-\tau,\;\tau]$, one can combine the aforementioned three cases, and express compactly the optimum as follows
 \begin{equation*}
  \mathbf{w}^\ast:={\rm proj}_{\tau}({c}/{\|\mathbf{a}\|_2^2})\cdot\mathbf{a}
 \end{equation*}
which concludes the proof of the proposition.

	\end{IEEEproof}

\bibliographystyle{IEEEtran}
\bibliography{myabrv,power}

\end{document}